\documentclass[aps,prd,preprintnumbers,nofootinbib,showpacs,onecolumn,superscriptaddress]{revtex4}%

\usepackage[dvips,dvipdfmx]{graphicx}
\usepackage{bm,latexsym,amsmath,amssymb,amsfonts,mathrsfs}
\usepackage{color}

\begin{document}

\vspace{2cm}
\title{\large Late-time quantum radiation by a uniformly accelerated detector in de Sitter spacetime}

\author{Shumpei Yamaguchi}
\affiliation{Department of Physics, Graduate School of Science, Hiroshima University,
  Higashi-Hiroshima 739-8526, Japan}
\author{Rumi Tatsukawa} 
\affiliation{Department of Physics, Graduate School of Science, Hiroshima University,
  Higashi-Hiroshima 739-8526, Japan}
\author{Shih-Yuin Lin}
\email{sylin@cc.ncue.edu.tw}
\affiliation{Department of Physics, National Changhua University of Education, Changhua 50007, Taiwan}
\author{Kazuhiro Yamamoto}
\email{kazuhiro@hiroshima-u.ac.jp}
\affiliation{Department of Physics, Graduate School of Science, Hiroshima University,
  Higashi-Hiroshima 739-8526, Japan}

\begin{abstract}
We investigate the quantum radiation emitted by a uniformly accelerated
Unruh-DeWitt detector in de Sitter spacetime.
We find that there exists a non-vanishing quantum radiation at late times 
in the radiation zone of the conformally flat coordinates, which cover the region
behind the cosmological horizon for the accelerated detector.
The theoretical structure of producing the late-time quantum radiation is
similar to that of the same model in Minkowski spacetime: 
it comes from a nonlocal correlation of the quantum field 
in the Bunch-Davies vacuum state, which can be traced back to the
entanglement between the field modes defined in different regions in de Sitter spacetime.
\end{abstract} 


\maketitle

\def\bar{\overline}
\def\x{{z}}
\def\L{L}
\def\bmx{{\bm x}}
\def\bmy{{\bm y}}
\def\bmp{{\bm p}}
\def\bmk{{\bm k}}
\def\deltatau{{\Delta\tau}}
\def\barphi{{\chi}}
\def\barpi{{\pi_\chi}}

\section{Introduction}

A localized object linearly and uniformly accelerated in the Minkowski vacuum would experience a thermal noise at a temperature proportional to its proper acceleration \cite{Unruh, ref-4}. This effect can be obtained straightforwardly by well established theories, but is very difficult to be detected in laboratories since the overall factor in the above ``Unruh temperature" of the noise is so tiny that the proper acceleration of the localized object has to be very large for detectable signals. One possibility in experiment may be using intense laser fields to accelerate charged particles, as proposed in Refs.~\cite{ChenTajima, Schutzhold, Schutzhold2, ELI,IYZ}, where the main observable is the quantum correction by the above ``Unruh effect" to the classical radiation of the charges accelerated by laser fields.
A major concern on those proposals is that quantum fluctuations of the field driving the charge and the response of the driven charge are perfectly coherent under equilibrium conditions at late times. It has been shown that in (1+1) dimensional Minkowski spacetime, indeed, a uniformly accelerated Unruh-DeWitt detector with derivative coupling has no quantum radiation under equilibrium conditions because of quantum interference \cite{Raine,Raval,Ford}. 
It seems that the only chance of detecting something would be in non-equilibrium situations
\footnote{The laser field is never static, and the acceleration of the driven charge is not uniform. In such non-equilibrium conditions the quantum radiation of an Unruh-DeWitt detector is indeed non-zero \cite{Lin16, Lin17}. However, non-equilibrium conditions are not the main point of this paper, see below.}.

While the equilibrium argument for the vanishing late-time quantum radiation in (1+1)D sounds convincing, it turns out that a uniformly accelerated Unruh-DeWitt detector in (3+1)D Minkowski spacetime still emits nonzero quantum radiation after relaxation, though the radiation is not originated from the thermal radiance that the detector experiences as in the Unruh effect \cite{LH2006}. This implies that more physics than equilibrium conditions have been involved in the late-time quantum radiation associated with the Unruh effect. Recently the authors of Refs.~\cite{IOTYZ,ITUY,HIUY} pointed out that the quantum radiation in similar models is in fact originated from a nonlocal property of the Minkowski vacuum state of a field, which induces the Unruh effect at the same time. The quantum radiation is nonvanishing in general, and the vanishing result obtained earlier in (1+1) dimensions is a special case due to the property of the field theory in two dimensional Minkowski spacetime.

For a uniformly accelerating localized object, the Rindler coordinates would be its natural reference frame \cite{BiD}. The Rindler coordinates only cover a part of Minkowski spacetime, called the R-wedge, which is causally disconnected with its conjugate Rindler wedge constructed on its opposite side (the L-wedge). 
It is well known that the Minkowski vacuum state of a field can be described by an entangled state of the Fock states of the field modes defined separately in those two Rindler wedges \cite{Unruh,UnruhWald,Higuchi}. The entanglement between the field modes in the two wedges can be understood as the origin of quantum radiation produced by a uniformly accelerated Unruh-DeWitt detector under equilibrium conditions
\cite{IOTYZ,ITUY,HIUY}.

The above insight from the simple Unruh-DeWitt detector models also applies to the quantum radiation by a relativistic charged particle in uniform acceleration coupled to vacuum fluctuations \cite{OYZ15,OYZ16}. One may wonder if this observation is general in any spacetime with similar properties such as the existence of different time-like Killing vectors in different coordinates and so different stationary states of the field, and that some coordinates only cover part of the space and others are global, etc. 

De Sitter spacetime is maximally symmetric and sufficiently simple, while it is important in describing the inflationary phase of the early universe as well as the accelerating expansion of our Universe at the present epoch. Properties of the quantum fields in de Sitter spacetime have recently attracted much attention in relativistic quantum information and cosmology \cite{KukitaNambu,Rotondo,KukitaNambu2,  MatsumuraNambu, MaldacenaPimentel, Kanno, KST}. It is well known that vacuum fluctuations of quantum fields in de Sitter spacetime exhibit thermal properties \cite{GH,BiD,BD,ref-4,MTY,IYZ2013} characterized by the Gibbons-Hawking temperature $T_{\rm GH}=H/2\pi$, where $H$ is the Hubble parameter \cite{GH}. It is also known that an Unruh-DeWitt detector uniformly accelerated in de Sitter spacetime would experience thermal fluctuations at a temperature $T=\sqrt{H^2+A^2}/2\pi$ with the proper acceleration $A$ in the weak coupling limit, reflecting the combination of the Unruh effect and the Gibbons-Hawking effect \cite{UdeS}.
The quantum radiation produced by an Unruh-DeWitt detector in de Sitter spacetime, including both the cases of inertial and uniformly accelerated detectors, would be a good starting point for our study beyond Minkowski spacetime.

Nevertheless, not every nice model in Minkowski spacetime behaves normally in De Sitter spacetime.  
For example, if a massless scalar field is minimally coupled to the scalar curvature of the background de Sitter spacetime, a classical particle coupled to the field with a monopole interaction can radiate even in inertial motion, as shown in Ref. \cite{Burko}, where the authors found that in response to this classical radiation the mass of the particle decreases in time and never reach an equilibrium state. 
Fortunately, in Ref.~\cite{Akhmedov2010} the authors investigated various models and found that in de Sitter spacetime, inertial charges moving in (i) electromagnetic fields, or (ii) a massless scalar field conformally coupled to the scalar curvature, do not emit classical radiation
\footnote{This does not imply that the models with the inertial charges radiating in de Sitter spacetime are ill-behaved, see 
\cite{Akhmedov2012}.}. The charged particles or detectors interacting with the scalar field (ii) possess the simplest possibilities for considering the quantum radiation under equilibrium conditions in de Sitter spacetime.

The photon emission of a spinless electric charge in de Sitter spacetime has been calculated in Refs. \cite{Blaga2015, BB16}, along the same line of Refs. \cite{NSY06, KNY11} using the time-dependent perturbation theory in the in-out formalism. However, time dependent perturbation theory gives the radiation rates in transient rather than those at late times. For our purpose,
we have to go beyond the perturbation theory. 
In the present paper, therefore, we will start with an exactly solvable model for an Unruh-DeWitt detector in (3+1) dimensional de Sitter spacetime, with the internal harmonic oscillator minimally coupled to a massless scalar field while the quantum field conformally coupled to the scalar curvature of the classical spacetime. 
We will demonstrate that an Unruh-DeWitt detector in uniform acceleration produces late-time quantum radiation, while an inertial detector along a geodesic does not. 
The origin of this nonvanishing late-time quantum radiation under equilibrium conditions 
can be interpreted as a consequence of the nonlocal correlation of the vacuum state in de Sitter spacetime.
This quantum radiation is different from the one emitted by the Unruh-DeWitt detector in oscillatory or circular motion in the late-time non-equilibrium stationary state \cite{Lin16, Lin17, BL83, BL87, Un98, AS07}.

This paper is organized as follows. In section \ref{model}, we introduce the model consisting of an Unruh-DeWitt detector and a massless scalar field in (3+1) dimensional de Sitter spacetime. 
We assume that the detector follows a trajectory in uniformly acceleration. Then in section \ref{solution}, the Heisenberg equations are solved under the stationary-state condition. We compute the two-point function of the field in section \ref{2pCorr} and find that a cancellation similar to the models in Minkowski spacetime also occurs in de Sitter spacetime.
We then compute the expectation value of the stress-energy tensor and the energy radiation rate in section \ref{LTRad} from the remaining terms of the two-point function of the field. 
A discussion on (1+1)D case is given in section \ref{discuss2D} before our conclusion in section \ref{conclude}. Throughout the paper, we use the unit $c=\hbar =1$.

\section{Detector-Field model in De Sitter spacetime}
\label{model}
\subsection{Equations of motion}
In the flat 
coordinates of de Sitter spacetime, the line element is written as
\begin{eqnarray}
  ds^2=dt^2-a^2(t)d{\bf x}^2,
  \label{dsdta}
\end{eqnarray}
where the scale factor is given by $a(t)=e^{Ht}$ with a constant $H$ and the cosmic
time $t$. Here the spatial section of the line element is described by the
comoving coordinate $d{\bf x}^2=\delta_{ij}dx^i dx^j$, for which we use the notation
$(x^1,x^2,x^3)=(x,\bf x_\perp)$.

In de Sitter spacetime, we consider the model consisting of a detector $Q$
and a massless scalar field $\phi$ with the action
\begin{equation}
  S=S_{0}(Q)+S_{int}(Q,\phi)+S_0(\phi),
\label{action}
\end{equation}
with
\begin{eqnarray}
 &&S_{0}(Q)=\int d\tau \frac{1}{2}\Bigl((\partial_{\tau} Q)^2-\Omega_{0}^2Q^2\Bigr),
\label{1-1}
\\
 &&S_0(\phi) =\int {d^4x} \sqrt{-g}\   \frac{1}{2} 
     \left\{ g^{\mu\nu} \partial_\mu\phi  \partial_\nu\phi 
                  -\xi R\phi^2 \right\} ,
\label{1-2}
\\
 &&S_{int}(Q,\phi) 
=  \lambda\int d\tau \int d^4x Q(\tau) \phi(x) \delta_D^{(4)}(x- z(\tau))
=  \lambda\int d\tau Q(\tau) \phi(\x^\mu(\tau)) ,
\label{1-3}
\end{eqnarray}
where $z^\mu(\tau)$ denotes the detector's trajectory parametrized by its proper time $\tau$ 
in the coordinates (\ref{dsdta}), and
$\xi$ is the non-minimal coupling constant to the spacetime curvature.
In the present paper we adopt the conformal coupling $\xi=1/6$.

The Heisenberg equations of motion give
\begin{eqnarray}
  &&  \partial_{\tau}^2 \hat Q+\Omega_{0}^2\hat Q={\lambda}\hat \phi(\x(\tau))
\\
&& \frac{1}{a^3}\partial_t(a^3 \partial_t \hat \phi)- \frac{\triangle \hat\phi}{a^2}+\xi R\hat\phi
    =\frac{\lambda}{a^3}\int d\tau \hat Q(\tau) \delta^4_D(x-z(\tau)).
\end{eqnarray}
Here we used the coordinates $(t,x,\bf x_\perp)$ of Eq.(\ref{dsdta}),
then $\delta^4_D(x-z(\tau))$ should be understood as
\begin{eqnarray}
  \delta^4_D(x-z(\tau))=\delta_D(t-t(\tau))\delta^3({\bf x}-{\bf z}(\tau)). 
\end{eqnarray}

Introducing the conformal time $\eta (-\infty<\eta<0)$ by
\begin{eqnarray}
  e^{Ht}=-{1\over H\eta}\equiv a(\eta), \label{t2eta}
\end{eqnarray}
the line element Eq.~(\ref{dsdta}) can be rewritten as
\begin{eqnarray}
  ds^2=a(\eta)^2(d\eta^2-d{\bf x}^2)=a(\eta)^2 \eta_{\mu\nu}d\bar x^\mu d\bar x^\mu ,
\label{dsdta2}
\end{eqnarray}
Let us introduce the field $\chi(x)$ by rescaling the field $\phi$, 
\begin{eqnarray}
  \hat\phi={\hat{\barphi} \over a}.
\end{eqnarray}
Then the equations of motion reduce to
\begin{eqnarray}
&&  \partial_{\tau}^2 \hat Q+\Omega_{0}^2\hat Q={\lambda}
{\hat {\barphi}(\bar\x(\tau))\over a(t(\tau))},
\label{EOMa}
\\
&&  \partial_\eta^2\hat{\barphi}- {\triangle \hat{\barphi}}
    ={\lambda} 
\int {d\tau\over a(\tau)} \hat Q(\tau) \delta^4_D(\bar x-\bar z(\tau)),
\label{EOMb}
\end{eqnarray}
where $\bar x^\mu =(\eta,{\bf x})=(\eta,x,x_\perp^1,x_\perp^2)$, and $\delta^4_D(\bar x-\bar z(\tau))$ in Eq.~(\ref{EOMb}) means
\begin{eqnarray}
  \delta^4_D(\bar x-\bar z(\tau))=\delta_D(\eta-\eta(\tau))\delta^3({\bf x}-{\bf z}(\tau)),
\end{eqnarray}
where we have used the relation $\delta_D(t-t(\tau))=\delta_D(\eta-\eta(\tau))/a(\tau)$.
Note that $\partial_\eta^2-\triangle\equiv\bar \square$ in Eq.~(\ref{EOMb})
is the  d'Alembert operator with respect to the (conformally
transformed) flat spacetime coordinate metric $\eta_{\mu\nu}$.

\subsection{Uniform acceleration in de Sitter space}

Below we present the detector's trajectory in uniform acceleration
in de Sitter spacetime, which is denoted by
$z^\mu(\tau)$ in the coordinates of Eq.~(\ref{dsdta}), or equivalently by 
$\bar z^\mu(\tau)$ with the coordinates of Eq.~(\ref{dsdta2}).
Consider a detector that moves along the trajectory given in \cite{UdeS},
\begin{eqnarray}
x e^{Ht}=K, ~~ x^2 = x^3=0,
\end{eqnarray}
where $K$ is a constant, in the coordinates of (\ref{dsdta}). Then one has $d\tau^2=dt^2 + a(t)^2 dx^2=(1-K^2 H^2)dt^2$, which implies
\begin{equation}
  t={\tau \over \sqrt{1-H^2K^2}},
\end{equation}
if we choose $\tau=0$ when $t=0$. So the four velocity of this detector is
\begin{equation}
  v^\mu=\frac{dz^\mu}{d\tau} = \frac{1}{\sqrt{1-H^2K^2}}\frac{d}{dt} (t, Ke^{-Ht},0,0)=
	{1\over \sqrt{1-H^2K^2}}(1,-HK e^{-Ht},0,0),
\end{equation}
which gives $v^\mu v_\mu=1$. The four acceleration vector of the trajectory now reads
\begin{equation}
  a^\mu= \frac{dv^\mu}{d\tau} 
  ={1\over {1-H^2K^2}}(H^3K^2,-H^2 Ke^{-Ht},0,0),
\end{equation}
which satisfies 
\begin{eqnarray}
&& a^\mu a_\mu=-{H^4 K^2\over 1-H^2K^2}\equiv -A^2,
\label{defA}
\end{eqnarray}
where the proper acceleration $A$ is a constant of time.
In this sense $z^\mu(\tau)$ is under uniform acceleration.

In the static coordinates for de Sitter space, 
the worldline $z^\mu$ is actually a timelike curve with a positive constant $r$, 
which is the distance from the detector to the observer at the origin ($r=0$).

\begin{figure}[t]
\includegraphics[width=0.45\textwidth]{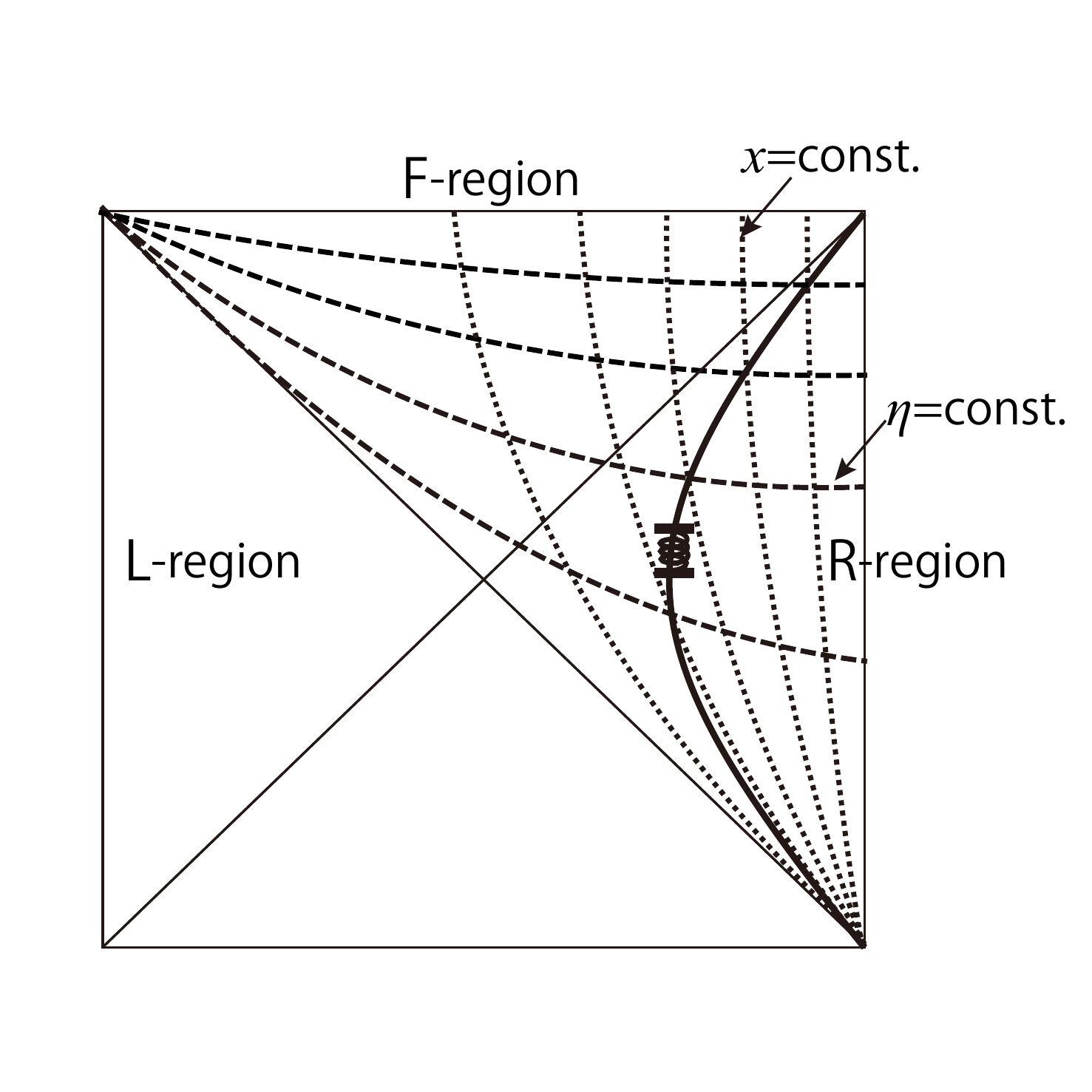} 
\caption{Penrose diagram of the conformally flat coordinates for de Sitter space. 
The bold solid curve represents the worldline of the uniformly accelerated detector, which is a 
straight line in the conformally flat coordinates.}
\label{fig:dSflat}
\end{figure}

Writing
\begin{eqnarray}
  a(t(\tau))=e^{Ht(\tau)}=e^{H\tau/\sqrt{1-H^2K^2}}\equiv e^{\alpha \tau}, \label{t2tau}
\end{eqnarray}
where we defined
\begin{eqnarray}
  \alpha={H\over \sqrt{1-H^2K^2}} = \sqrt{A^2+H^2}, \label{defAlpha}
\end{eqnarray}
then the trajectory of the uniformly accelerated detector can be parametrized by its proper time in the coordinates (\ref{dsdta}) as 
\begin{eqnarray}
	z^\mu(\tau)=(\tau/\sqrt{1-H^2K^2},Ke^{-H\tau/\sqrt{1-H^2K^2}},0,0), \label{zoftau}
\end{eqnarray}
or in the conformally flat coordinates (\ref{dsdta2}) as
\begin{eqnarray}
  \bar z^\mu=\left(-{e^{-\alpha\tau}\over H},Ke^{-\alpha\tau},0,0\right), \label{zbartau}
\end{eqnarray}
which is a straight line satisfying $\bar z^\mu \bar z_\mu=\alpha^{-2}$ in the chart 
(see figure \ref{fig:dSflat}).

We note that the hypersurfaces $\eta^2-\rho^2=0$, $\rho\equiv |{\bf x}| = \sqrt{x^2+|{\bf x}_\perp|^2}$ 
are the horizons for a local observer at rest along the worldline $(\eta,0,0,0)$ 
in the conformally flat coordinates (\ref{dsdta2}) (which are the $r=1/H$ hypersurfaces
in the static coordinates for de Sitter space).
Analogous to the Rindler wedges in Minkowski space, below we call the regions with 
$\eta^2-\rho^2> 0$ and $\eta^2-\rho^2 < 0$ as the R-region and F-region, respectively 
(see figure \ref{fig:dSflat}).
The whole worldline of the uniformly accelerated detector (\ref{zbartau}) is inside
the R-region and has a
two-way causal connection with the inertial observer resting at the spatial origin.

\section{Solution for the Detector-Field Equations}
\label{solution}
\subsection{Detector's equation of motion with radiation reaction}

Eq. (\ref{EOMb}) can be formally solved to obtain
\begin{eqnarray}
  \hat\barphi=\hat\chi_h(\bar x)+\hat\chi_{inh}(\bar x),
  \label{jbcc}
\end{eqnarray}
where $\chi_h$ is the homogeneous solution, which is given by the free quantized field,
and $\chi_{inh}$ is the inhomogeneous solution given by 
\begin{eqnarray}
  \hat\chi_{inh}(\bar x)
  =\lambda\int d^4\bar x' \bar G_R(\bar x,\bar x')\int_{\tau_0}^\infty {d\tau \over a(\tau)} \hat Q(\tau) \delta^4_D(\bar x'-\bar z(\tau)),
  \label{chiinh}
\end{eqnarray}
in the conformally flat coordinates, assuming the coupling is switched-on at $\tau=\tau_0$.
Here the retarded Green function
\begin{eqnarray}
  \bar G_{R}(\bar x,\bar x')\equiv
  {1\over 4\pi}\delta_D\bigl(\bar\sigma\bigr)\theta(\eta-\eta'), \label{Gret}
\end{eqnarray}
with Synge's world function  
\begin{eqnarray}
  \bar \sigma(\bar x,\bar x')={1\over 2}\eta_{\mu\nu}(\bar x^\mu-\bar x'{}^\mu)
  (\bar x^\nu-\bar x'{}^\nu)
       ={1\over 2}\left((\eta-\eta')^2-|{\bf x}-{\bf x}'|^2\right),
\end{eqnarray}
is simply identical to the one in Minkowski space
by virtue of the conformal coupling to the curvature.
The retarded Green's function can be more complicated in the cases with $\xi\not= 1/6$
(e.g., for the minimal coupling ($\xi=0$), see Ref.\cite{Burko}).

Inserting the worldline (\ref{zoftau}) into (\ref{chiinh}) while
taking into account the relation between the cosmic time and the conformal time
with (\ref{t2eta}) and (\ref{t2tau}), we have
\begin{eqnarray}
  \hat\chi_{inh}(\bar x)&=&\lambda\int_{\zeta_0}^0 {d\zeta}\sqrt{1-H^2K^2}\bar G_{R}(\bar x,\bar z(\zeta)) \hat Q(\tau(\zeta)),
	\label{chiinheta}
\end{eqnarray}
with
\begin{eqnarray}
  \zeta=-{1\over H} e^{-\alpha\tau} \label{zeta2tau}
\end{eqnarray}
and $\zeta_0 = \zeta(\tau_0)$.
In solving (\ref{EOMa}), one needs to know $\hat\chi_{inh}(\bar z)$, but 
$\hat\chi_{inh}(\bar x) \propto |\bar{\bf x}-\bar{\bf z}(\zeta)|^{-1}$ diverges in the coincidence limit $\bar x \to \bar z(\zeta)$. 
To control the divergence, we may use the regularized retarded Green function \cite{LH2006,JohnsonHu},
\begin{eqnarray}
  \bar G_{R}^{\bar{\Lambda}}(\bar x,\bar x')
   = {1\over4\pi}\sqrt{8\over \pi}\bar\Lambda^2e^{-2\bar\Lambda^4\bar \sigma^2}\theta(\eta-\eta') 
	\stackrel{\bar{\Lambda}\to\infty}{\rightarrow}	\bar G_{R}(\bar x,\bar x'),
\label{regularization}
\end{eqnarray}
where \textcolor{black}{we introduce the UV regulator $\bar\Lambda$ in the conformally transformed flat spacetime of the coordinate $\bar x^\mu$ following the philosophy of the effective field theory.}
Then we have the local expansion
\begin{eqnarray}
 &&\lambda\int_{\zeta_{0}}^0 {d\zeta'}\sqrt{1-H^2K^2}\bar G_{R}(\bar{z}(\zeta),\bar z(\zeta')) \hat Q(\tau(\zeta'))
  \simeq{\lambda \over 4\pi}\left(\bar{\Lambda}\beta-{1 \over {\sqrt{1-H^2K^2}}}{d \over d\zeta}
  +{\cal O}(\bar \Lambda^{-1})\right)\hat Q(\tau(\zeta)) \label{intLam}
\end{eqnarray}
where $\beta=2^{7/4}\Gamma(5/4)/\pi^{1/2}$, and we have used
\begin{eqnarray}
  {1\over 2}\eta_{\mu\nu}(\bar x^\mu(\zeta)-\bar x^\mu{}'(\zeta'))(\bar x^\nu(\zeta)-\bar x^\nu{}'(\zeta'))=
       {1\over 2}(1-H^2K^2)(\zeta-\zeta')^2.
\end{eqnarray}
Substitute (\ref{jbcc}) and (\ref{chiinheta}) with (\ref{intLam}) into (\ref{EOMa}), we find that Eq. (\ref{EOMa}) reduces to 
\begin{eqnarray}
&&\biggl({d^2\over d\tau^2}+2\gamma 
{d\over d\tau}+\Omega^2\biggr)
\hat Q(\tau)={\lambda\over a(t(\tau))}\hat\chi_h(\bar z(\tau)),
\label{joj}
\end{eqnarray}
in the large $\Lambda$ limit. Here $\gamma\equiv \lambda^2/8\pi$, and we have
introduced $\bar \Lambda/a=\Lambda$ and a renormalized frequency,
\begin{eqnarray}
  \Omega^2=\Omega_{0}^2-{\lambda^2\Lambda\beta \over 4\pi}.
\end{eqnarray}
We note that 
$\Lambda$ is the UV cutoff with respect to the proper time of the detector, 
\textcolor{black}{corresponding to the shortest timescale of the detector and the highest frequency of the field modes that the detector can sense. So $\Lambda$ should be taken as a constant in the physical coordinate of the detector, while $\bar{\Lambda} = a(\eta)\Lambda$ 
would depend on the conformal time $\eta$, which is not directly measurable anyway. 
For the regulator $\bar{\Lambda}$ depending only on $\eta$ and independent of $\eta'$ in (\ref{regularization}), 
the expansion in (\ref{intLam}) has the same form as the one for a constant $\bar{\Lambda}$.}

In this paper, we are interested in the late-time behavior of the field ($\tau_0, \zeta_0 \to -\infty$).
The late-time steady state solution for Eq.~(\ref{joj}) is given by
\begin{eqnarray}
&&  \hat Q(\tau)={1\over 2\pi}\int_{-\infty}^\infty d\omega \tilde Q(\omega) e^{-i\omega\tau},
 \label{Qlate}
\end{eqnarray}
where 
\begin{eqnarray}
   h(\omega)&=&{1\over -\omega^2- 2 i\omega \gamma 
   +\Omega^2}, \label{suscept} \\
  \tilde Q(\omega)&=&\lambda h(\omega)\delta \tilde \varphi(\omega), 
\end{eqnarray}
with the Fourier-transformed homogeneous solution for the field $\delta\tilde\varphi$ with respect to $\tau$ defined by
\begin{eqnarray}
&&  {\hat\chi_h(\bar z(\tau))\over a(t(\tau))}= {1\over 2\pi}\int d\omega \delta\tilde \varphi(\omega) e^{-i\omega\tau}.
\label{deltaphi}
\end{eqnarray}
Then the inhomogeneous solution (\ref{chiinh}) can be written as
\begin{eqnarray}
  &&{\chi_{inh}(\bar x)}= \frac{\lambda^2}{2\pi}\int_{-\infty}^\infty d\omega h(\omega)\delta\tilde\varphi(\omega) \int_{-\infty}^\infty 
	{d\tau \over a(t(\tau))} e^{-i\omega\tau} \bar G_R(\bar x-\bar z(\eta(\tau)))
\end{eqnarray}
at late times.

\section{two-point function of Field}
\label{2pCorr}

Now, we consider the two-point function defined by
\begin{eqnarray}
  \langle \hat\chi(\bar x)\hat\chi(\bar x')\rangle-\langle \hat\chi_h(\bar x)\hat\chi_h(\bar x')\rangle
  =
  \langle \hat\chi_{inh}(\bar x)\hat\chi_h(\bar x')\rangle+\langle
  \hat\chi_h(\bar x)\hat\chi_{inh}(\bar x')\rangle
  +\langle \hat\chi_{inh}(\bar x)\hat\chi_{inh}(\bar x')\rangle,
  \label{it}
\end{eqnarray}
assuming that the quantum field is initially in the Bunch-Davies vacuum \cite{BiD, BD}.

\subsection{Inhomogeneous Term $\langle\chi_{inh}(\bar x)\chi_{inh}(\bar x')\rangle$}
  The correlation function of the inhomogeneous solutions, namely, the last term of Eq.~(\ref{it}), in the Bunch-Davies vacuum,
can be worked out straightforwardly:
\begin{eqnarray}
\langle\hat\chi_{inh}(\bar x)\hat\chi_{inh}(\bar x')\rangle 
&=& 
 {-i\lambda^2 H^2\over (4\pi)^2\alpha^2{\cal R}(\bar x){\cal R}(\bar x')}
\int{d\omega \over 2\pi}\left[h(\omega)-h(-\omega)\right]{e^{-i\omega(\tau^F_-(\bar x)-\tau^F_-(\bar x'))} \over 1-e^{-{2\pi\omega /\alpha}}}.
\label{chiichii}
\end{eqnarray}
Here we have used the identity
\begin{eqnarray}
  h(\omega)-h(-\omega)= 4i\gamma\omega 
	|h(\omega)|^2,
  \label{FDR}
\end{eqnarray}
which has the form of the fluctuation-dissipation theorem.
One can read off the Gibbons-Hawking-Unruh temperature $T_\alpha = \alpha/(2\pi k^{}_B)$ from the Planck factor 
in the integrand of the above result. Since we have taken the frequency $\omega$ with respect to the proper time $\tau$
of the uniformly accelerated detector in (\ref{Qlate}), $T_\alpha$ may be interpreted as the temperature experienced by the 
detector in the Bunch-Davies vacuum in the weak-coupling limit.

\subsection{Interference Term $\langle\hat\chi_h(\bar x)\hat\chi_{\rm inh}(\bar x')\rangle$}
The first term on the right hand side of Eq.~(\ref{it}) can be expressed as
\begin{eqnarray}
&&\langle\hat\chi_h(\bar x)\hat\chi_{\rm inh}(\bar x')\rangle=\lambda^2 \int d\tau {1\over a(t(\tau))}{1\over 2\pi} 
\int d\omega h(\omega)e^{-i\omega\tau} \bar G_R(\bar x'-\bar z(\zeta)) \langle\hat\chi_h(\bar x)\delta\tilde\varphi\rangle,
\end{eqnarray}
in which, by (\ref{deltaphi}),
\begin{eqnarray}
  \langle\hat\chi_h(\bar x)\delta\tilde\varphi\rangle=
  \int_{-\infty}^\infty d\tau {\langle\hat\chi_h(\bar x)\hat\chi_h(\bar z(\tau))\rangle \over a(t(\tau))} e^{i\omega \tau}
  \label{chivarphi}
\end{eqnarray}
with the positive frequency Wightman function
\begin{eqnarray}
&&\langle\hat\chi_h(\bar x)\hat\chi_h(\bar z(\tau))\rangle=
-{1\over 4\pi^2[(\eta-\zeta(\tau-i\varepsilon))^2-(x+KH\zeta(\tau-i\varepsilon))^2-x_\perp^2]} 
= -\frac{1}{8\pi^2\bar\sigma(\bar x, \bar z(\tau))}
\end{eqnarray}
($\varepsilon\to 0+$) in the Bunch-Davies vacuum, which is equivalent to Minkowski space Wightman function.
The poles of the above integrand on the complex plane of $\tau$ will be found by solving $\zeta(\tau)$ in
\begin{eqnarray}
  0 &=& 2\bar\sigma(\bar x, \bar z(\tau)) = (\eta-\zeta(\tau-i\varepsilon))^2-(x+KH\zeta(\tau-i\varepsilon))^2-x_\perp^2 \nonumber\\
	  &=& (1-K^2H^2)\zeta^2 -2(\eta+K H x)\zeta + \eta^2-\rho^2. 
\end{eqnarray}
The poles will depend on the sign of $\eta^2-\rho^2$.
Below we discuss the cases with $\bar x^\mu$ in the R-region ($\eta^2-\rho^2 >0$) and 
in the F-region ($\eta^2-\rho^2<0$) separately.

For $\bar x^\mu$ in the R-region, $\eta^2-\rho^2>0$, 
and the poles in the complex $\tau$ plane are located at
\begin{eqnarray}
\tau=\tau_\pm^R+i2\pi n/\alpha-i\varepsilon, ~~~(n=0,\pm1,\pm2,\cdots), 
\end{eqnarray}
where $\tau_\pm^R$ are the principal values defined by
\begin{eqnarray}
  \tau^R_\pm(\bar x) \equiv -\frac{1}{\alpha} \ln\left[ 
	{\alpha^2\over H}\left(-\eta-HKx\mp {\cal R}(\bar x)\right)\right]
\end{eqnarray}
with 
\begin{equation}
  {\cal R}(\bar x) \equiv \sqrt{(HK\eta+x)^2+(1-H^2K^2)x_\perp^2}.
\end{equation}
In fact, ${\cal R}$ is the counterpart of the Minkowski retarded distance
\cite{OLMH12} in the conformally flat coordinates for de Sitter space, namely,
\begin{equation}
  \left| \frac{d}{d\eta}\bar\sigma(\bar x, \bar z(\tau)) \right|_{\tau=\tau_-^{R}} = {\cal R}(\bar x).
\end{equation}

Applying the residue theorem to the integration, we obtain 
\begin{equation}
 \int_{-\infty}^\infty d\tau{e^{i\omega\tau}\over e^{\alpha\tau}[(\eta-\zeta(\tau)-i\varepsilon)^2-(x+KH\zeta(\tau))^2-x_\perp^2]}
= {-i\pi H\over \alpha {\cal R}(\bar x)}
\left({e^{i\omega\tau^R_-}\over e^{2\pi\omega/\alpha}-1}
-{e^{i\omega\tau^R_+}\over e^{2\pi\omega/\alpha}-1}\right).
\end{equation}

For $\bar x^\mu$ in the F-region, $\eta^2-\rho^2<0$, 
the poles are located at 
\begin{eqnarray}
  \tau=\tau_-^F+i2\pi n/\alpha-i\varepsilon,
	\hspace{.5cm} {\rm and} \hspace{.5cm}
  ~\tau^F_+ +i\pi (2n-1)/\alpha,
  ~~(n=0,\pm1,\pm2,\cdots), 
\end{eqnarray}
where $\tau_\pm^F$ are defined by
\begin{eqnarray}
  \tau^F_\pm(\bar x) &\equiv&-\frac{1}{\alpha}\ln \left[{\alpha^2\over H}\left(\pm(\eta+HKx)+
  {\cal R}(\bar x)\right)\right].
\label{taufpm}
\end{eqnarray}
Still, ${\cal R}(\bar x)=\left| \frac{d}{d\eta}\bar\sigma(\bar x, \bar z)\right|_{\tau=\tau_-^F}$ is the counterpart of the
Minkowski retarded distance here.
By the residue theorem, we obtain
\begin{eqnarray}
&&\int_{-\infty}^\infty d\tau{e^{i\omega\tau}\over e^{\alpha\tau}[(\eta-\zeta(\tau)-i\varepsilon)^2-(x+KH\zeta(\tau))^2-x_\perp^2]}=
	{-i\pi H\over \alpha {\cal R}(\bar x)}
\left(
{e^{i\omega\tau^F_-}\over e^{2\pi\omega/\alpha}-1}
-{e^{\pi\omega/\alpha} e^{i\omega\tau^F_+}\over e^{2\pi\omega/\alpha}-1}\right).
\end{eqnarray}

\subsection{Subtracted Two-Point Function of Field in F-Region}
In a similar way to the case of an accelerated detector in Minkowski spacetime
\cite{IOTYZ,ITUY,HIUY}, we will show that the inhomogeneous term cancels out
by the interference term. Since we are interested in the quantum radiation,
which is evaluated in the F-region, we here consider the case
$\eta^2-\rho^2<0$. In this case, summarizing the result in
the previous section, we have
\begin{eqnarray}
  \langle\hat\chi_{h}(\bar x)\hat\chi_{inh}(\bar x')\rangle
  &=& {-i\lambda^2 H^2\over (4\pi)^2\alpha^2{\cal R}(\bar x){\cal R}(\bar x')}
	\int {d\omega \over 2\pi}
  h(\omega)\left({e^{\pi\omega/\alpha}e^{i\omega(\tau^F_+(\bar x)-\tau_-^F(\bar x'))}\over e^{2\pi\omega/\alpha}-1}
  -  {e^{i\omega(\tau_-^F(\bar x)-\tau_-^F(\bar x'))}\over e^{2\pi\omega/\alpha}-1}\right).
\end{eqnarray}
The $e^{i\omega(\tau_-^F(\bar x)-\tau_-^F(\bar x'))}$ term in the integrand of the above result and the conjugate term in 
$\langle\hat\chi_{inh}(\bar x)\hat\chi_{h}(\bar x')\rangle$
will cancel the whole inhomogeneous term $\langle\hat\chi_{inh}(\bar x)\hat\chi_{inh}(\bar x')\rangle$ in (\ref{chiichii}),
so the subtracted two-point function of the field reduces to 
\begin{eqnarray}
  &&\langle\hat\chi(\bar x)\hat\chi(\bar x')\rangle-\langle\hat\chi_h(\bar x)\hat\chi_h(\bar x')\rangle
={-i\lambda^2 H^2\over (4\pi)^2\alpha^2{\cal R}(\bar x){\cal R}(\bar x')}{\cal F}(\bar x,\bar x'),
\label{subtwo}
\end{eqnarray}
where we defined 
\begin{eqnarray}
{\cal F}(\bar x, \bar x')\equiv \int_{-\infty}^{+\infty} {d\omega \over 2\pi}
  {e^{\pi\omega/\alpha}\over e^{2\pi\omega/\alpha}-1}
  \left[h(\omega)e^{-i\omega(\tau^F_-(\bar x)-\tau^F_+(\bar x'))}-h(-\omega)e^{-i\omega(\tau^F_+(\bar x)-\tau^F_-(\bar x'))}\right]
	\label{Fdef}
\end{eqnarray}
for $\bar x$, $\bar x'$ are both in the F-region.
In terms of the original non-scaled scalar field $\phi$, the subtracted  two-point function of the field is simply
\begin{eqnarray}
&&\langle\hat\phi(\bar x)\hat\phi(\bar x')\rangle^{}_{ren}\equiv 
\langle\hat\phi(\bar x)\hat\phi(\bar x')\rangle-\langle\hat\phi_h(\bar x)\hat\phi_h(\bar x')\rangle
= {-i\lambda^2 H^2 {\cal F}(\bar x,\bar x')\over (4\pi)^2\alpha^2 a(\eta){\cal R}(\bar x)a(\eta'){\cal R}(\bar x')}.
\label{pppp}
\end{eqnarray}
We note that the fluctuation-dissipation relation in the form of Eq.~(\ref{FDR})
plays an important role in the cancellation of the inhomogeneous term.

The whole integrand of ${\cal F}$ in (\ref{Fdef}) converges as $\omega\to 0$. However, when we remove the square brackets and break the integrand into two terms in order to get a closed form of the integral, each term is singular at $\omega=0$ due to the Planck factor.
One can introduce a regulator $\epsilon\to 0+$ to the Planck factor for convenience to regularize the singularity, e.g. $1/(e^{2\pi\omega/\alpha}-1) \to 1/(e^{2\pi(\omega+i\epsilon)/\alpha}-1)$. The result should not depend on the sign of the regulator $\epsilon$, which can be checked by numerical method as well as an analytic way as shown in Appendix B of Ref.~\cite{OYZ15}.
In addition to the poles $\omega = in\alpha$, $n\in {\bf Z}$, for the Planck factor in Eq.(\ref{Fdef}), the function
$h(\omega)$ 
defined in (\ref{suscept})
has two poles at $\omega = \Omega_\pm$, where 
$\Omega_{\pm} \equiv {\gamma}\pm\sqrt{\gamma^2-\Omega^2}$ 
for  $\Omega< \gamma$ and $\Omega_{\pm} \equiv {\gamma}\pm i\sqrt{\Omega^2-\gamma^2}$ 
for  $\Omega > \gamma$. So ${\cal F}(\bar x,\bar x')$ in (\ref{Fdef}) can be expressed as
\begin{eqnarray}
  &&{\cal F}(\bar{x},\bar{x}') = 
	\frac{i}{2\pi(\Omega_+-\Omega_-)}\times \nonumber\\
	&& \left\{ \theta(\bar{\tau}_--\bar{\tau}'_+)\left[ \sum_{n=0}^\infty
	  \left( \frac{(-1)^n\alpha }{n\alpha-\Omega_-}-\frac{(-1)^n\alpha }{n\alpha-\Omega_+}\right)e^{-n\alpha(\bar{\tau}_--\bar{\tau}'_+)}
	+ \frac{\pi e^{-\Omega_-(\bar{\tau}_--\bar{\tau}'_+)}}{\sin(\pi\Omega_-/\alpha)}
	- \frac{\pi e^{-\Omega_+(\bar{\tau}_--\bar{\tau}'_+)}}{\sin(\pi\Omega_+/\alpha)}\right]\right.\nonumber\\
	&& + \theta(\bar{\tau}'_+-\bar{\tau}_-)\left[ \sum_{n=1}^\infty
	  \left( \frac{(-1)^n\alpha }{n\alpha+\Omega_-}-\frac{(-1)^n\alpha }{n\alpha+\Omega_+}\right)e^{-n\alpha(\bar{\tau}'_+-\bar{\tau}_-)}
		\right] \nonumber\\
  && + \theta(\bar{\tau}'_--\bar{\tau}_+)\left[ \sum_{n=1}^\infty
	  \left( \frac{(-1)^n\alpha }{n\alpha-\Omega_-}-\frac{(-1)^n\alpha }{n\alpha-\Omega_+}\right)e^{-n\alpha(\bar{\tau}'_--\bar{\tau}_+)}
	+ \frac{\pi e^{-\Omega_-(\bar{\tau}'_--\bar{\tau}_+)}}{\sin(\pi\Omega_-/\alpha)}
	- \frac{\pi e^{-\Omega_+(\bar{\tau}'_--\bar{\tau}_+)}}{\sin(\pi\Omega_+/\alpha)}\right]\nonumber\\
	&& \left. + \theta(\bar{\tau}_+-\bar{\tau}'_-)\left[ \sum_{n=0}^\infty
	  \left( \frac{(-1)^n\alpha }{n\alpha+\Omega_-}-\frac{(-1)^n\alpha }{n\alpha+\Omega_+}\right)e^{-n\alpha(\bar{\tau}_+-\bar{\tau}'_-)}
		\right]\right\},
\end{eqnarray}
where we denoted $\bar{\tau}_\pm \equiv \tau_\pm^F(\bar{x})$ and $\bar{\tau}'_\pm \equiv \tau_\pm^F(\bar{x}')$.

In next section we calculate the energy momentum tensor,
  where we introduce the symmetrized two-point correlator,
  $\bigl\{\langle \hat\phi(x)\hat\phi(x')\rangle^{}_{ren} +
    \langle\hat\phi(x')\hat\phi(x)\rangle^{}_{ren}\bigr\}/2$.
  Using the above result, we find that the symmetrized two-point correlator
  can be written as
  \begin{eqnarray}
  {1\over2}\left\{ \langle\hat\phi(x)\hat\phi(x')\rangle^{}_{ren} +
    \langle\hat\phi(x')\hat\phi(x)\rangle^{}_{ren}\right\}\
  =
  {-i\lambda^2 H^2 \left\{{\cal F}(\bar x,\bar x')+{\cal F}(\bar x',\bar x)\right\}
      \over 2(4\pi)^2\alpha^2 a(\eta){\cal R}(\bar x)a(\eta'){\cal R}(\bar x')}
    =
  {-i\lambda^2 H^2 \left\{{\cal G}(\bar x,\bar x')+{\cal G}(\bar x',\bar x)\right\}
      \over 2(4\pi)^2\alpha^2 a(\eta){\cal R}(\bar x)a(\eta'){\cal R}(\bar x')}
\end{eqnarray}  
where
\begin{eqnarray}
  {\cal G}(\bar x,\bar x')&=& -i\theta\left(\tau_-^F(\bar x')-\tau^F_+(\bar x)\right)\biggl[
  {1\over \Omega_+\Omega_-}{\alpha\over 2\pi}+f_-(\bar x,\bar x')+f_+(\bar x,\bar x')
	+\frac{1}{\alpha} \sum_{n=1}^\infty g_{\alpha,n}(\bar x,\bar x') \biggr] \nonumber\\
  & & +i\theta\left(\tau^F_+(\bar x)-\tau_-^F(\bar x')\right)\biggl[
    {1\over \Omega_+\Omega_-}{\alpha\over 2\pi}
    +\frac{1}{\alpha}\sum_{n=1}^\infty g_{-\alpha,n}(\bar x,\bar x')\biggr],
\end{eqnarray}
with
\begin{equation}
	f_\pm(x,x') \equiv {\pm e^{\Omega_\pm(\tau^F_+(x)-\tau^F_-(x'))}\over (\Omega_+-\Omega_-)\sin(\pi\Omega_\pm/\alpha)},\hspace{.5cm}
	g_{\kappa, n}(x,x')\equiv {\kappa^2\over \pi}{(-1)^n e^{n\kappa(\tau_+^F(x)-\tau^F_-(x'))}\over (\Omega_--n\kappa)(\Omega_+-n\kappa)}.
\label{fgdef}
\end{equation}

The subtracted two-point function in de Sitter spacetime (\ref{pppp})
has the properties similar
to that in Minkowski spacetime \cite{IYZ}. First, the inhomogeneous term is canceled
out by the interference term. Second, the factor
$e^{\pi\omega/\alpha}/(e^{2\pi\omega/\alpha}-1)$ appears in (\ref{Fdef}),
which is also a common property to the case in Minkowski spacetime.
As discussed in Refs.~\cite{IOTYZ,ITUY,HIUY},
the factor $e^{\pi\omega/\alpha}/(e^{2\pi\omega/\alpha}-1)$ comes from entanglement between
the field modes in different regions. The subtracted two-point function
reflects a nonlocal correlation of the quantum field 
in the Bunch-Davies vacuum state in de Sitter spacetime,
which can be traced back to the entanglement between the field modes
constructed in the partially covered regions in de Sitter spacetime.

\section{Radiation Rate} 
\label{LTRad}
In this section, we evaluate the radiation rate produced by an
  accelerated detector in de Sitter spacetime. 
  We work with the coordinate $(\eta,x,{\bf x}_\perp)$, which should be written
  ${\bar x}^\mu$.
  However, we omit the ``bar'', unless otherwise noted explicitly, for simplicity.

The classical energy momentum tensor of a scalar field with
non-minimal coupling term to the scalar curvature is 
\begin{eqnarray}
  T_{\mu\nu}=(1-2\xi)\nabla_\mu\phi\nabla_\nu\phi-2\xi\phi\nabla_\mu\nabla_\nu\phi+\left(2\xi-{1\over2}\right)g_{\mu\nu}\nabla^\alpha\phi\nabla_\alpha\phi
  +{\xi\over 2}g_{\mu\nu}\phi \nabla^\alpha\nabla_\alpha\phi
\end{eqnarray}
with $\xi=1/6$. In quantum theory, we promote the above field amplitude $\phi$ to operators $\hat\phi$ and calculate the renormalized expectation value of the stress-energy tensor operator:
\begin{eqnarray}
  \langle \hat T_{\mu\nu}(x)\rangle^{}_{ren} = \lim_{x'\to x}\left(\frac{2}{3}\nabla_\mu\nabla'_\nu-\frac{1}{3}\nabla'_\mu\nabla'_\nu -
	\frac{1}{6} g_{\mu\nu}\nabla^\alpha\nabla'_\alpha
  +\frac{1}{12} g_{\mu\nu} \nabla'^\alpha\nabla'_\alpha\right) \langle \hat\phi(x), \hat\phi(x')\rangle^{}_{ren}, \label{TuvNM}
\end{eqnarray}
where $\langle \hat\phi(x), \hat\phi(x')\rangle^{}_{ren} \equiv \langle (\hat\phi(x)\hat\phi(x') + \hat\phi(x')\hat\phi(x))\rangle^{}_{ren}/2$ is the symmetrized two-point correlator of the field, which is taken here because $\langle \hat T_{\mu\nu}(x)\rangle^{}_{ren}$ is measurable and so should be real. $\nabla$ and $\nabla'$ denote the covariant derivatives with respect of $x$ and $x'$, respectively,
such that
\begin{eqnarray}
&& \lim_{x'\rightarrow x}
  \nabla_{\mu}\nabla'_{\nu}\langle \hat\phi(x)\hat\phi(x')\rangle^{}_{ren}
  =\lim_{x'\rightarrow x}\left({\partial\over \partial x^\mu}
  {\partial\over \partial x'^{\nu}}\right)\langle \hat\phi(x)\hat\phi(x')\rangle^{}_{ren},
  \\
  && \lim_{x'\rightarrow x}
  \nabla'_{\mu}\nabla'_{\nu}\langle \hat\phi(x)\hat\phi(x')\rangle^{}_{ren}
  =\lim_{x'\rightarrow x}\left({\partial\over \partial x'^\mu}
  {\partial\over \partial x'^{\nu}}-\Gamma^\rho{}_{\mu\nu}(x')
	{\partial\over\partial x'^\rho}\right)\langle \hat\phi(x)\hat\phi(x')\rangle^{}_{ren}.
\end{eqnarray}

Given the worldline (\ref{zbartau}) and the corresponding four velocity $v^\mu(\tau) = \frac{d}{d\tau}z^\mu(\tau) = -\alpha z^\mu(\tau)$ 
for a uniformly accelerated detector in the conformally flat coordinates,  
we introduce the spacelike vector $u^\mu(\tau)$ by (cf. \cite{LH2006, Rohrlich})
\begin{eqnarray}
  u_\mu(\tau) u_\mu(\tau)=-1, ~~~~ u^\mu(\tau) v_\mu(\tau)=0
\end{eqnarray}
which implies
\begin{eqnarray}
  x^\mu-z^\mu(\tau_-(x))= 
	  \frac{\alpha}{H}{\cal R}(x) a(\tau_-(x)) \left[u^\mu(\tau_-(x))+v^\mu(\tau_-(x))\right],
\end{eqnarray}
such that $v_\mu(\tau_-)(x^\mu-z^\mu(\tau_-))= 
	  {\alpha}{\cal R} a(\tau_-)/H$, and $u_\mu(\tau_-)x^\mu =- 
	  {\alpha}{\cal R} a(\tau_-)/H$ because $u_\mu(\tau_-)z^\mu(\tau_-)
          \propto u_\mu(\tau_-)v^\mu(\tau_-)=0$.
 
By using the relations
\begin{eqnarray}
  &&{\partial\over \partial \bar x^\mu}
	{\cal R}(x)=
	{H^2\over \alpha^2{\cal R}(x)}\left(
  (\eta+HKx)(\delta^0{}_\mu+HK\delta^1{}_\mu)-{H^2\over \alpha^2 }\eta_{\mu\alpha}x^\alpha\right),
\\
&& {\partial \tau^F_{\mp}(x)\over \partial \bar x^\mu}
  ={\alpha\over H} e^{\alpha\tau^F_\mp(x)}
  \left\{\pm(\delta^0{}_\mu+HK\delta^1{}_\mu)-{1\over {\cal R}(x)}
  \left((\eta+HKx)(\delta^0{}_\mu+HK\delta^1{}_\mu)-{H^2\over \alpha^2}\eta_{\mu\alpha}x^\alpha\right)\right\},
\\
 &&e^{-\alpha\tau^F_+(x)}e^{-\alpha\tau^F_-(x)}=
  -\alpha^2(\eta^2-x^2-|{\bf x}_\perp|^2 ),
\end{eqnarray}
we end up with the following expression, after some tedious algebra, 
\begin{eqnarray}
  &&u^\mu(\tau^F_-(x))  v^\nu(\tau^F_-(x)) \langle T_{\mu\nu}(x)\rangle^{}_{ren}
   ={2\over 3}{\lambda^2\over (4\pi)^2}{H^2\over \alpha a^2(\eta){\cal R}^2(x)} {\cal J}(x),  \label{uvT}
\end{eqnarray}	
where we defined
\begin{eqnarray}
 && {\cal J}(x) \equiv
  {(r_- -\alpha^{-1})\over (2 r_- 
	-\alpha^{-1})^2}\times\nonumber\\
 &&\biggl[\theta(\tau^F_--\tau^F_+)
    \left\{
    ( r_- - \alpha^{-1})
    \biggl(
    \Omega_- f_-
    +\Omega_+ f_+
    +\sum_{n=1}^\infty n g_{\alpha,n}
		\biggr)
     - r_- 
		\biggl(
    {\Omega_-^2 f_-\over \alpha} 
    +{\Omega_+^2 f_+ \over \alpha} 
    +\sum_{n=1}^\infty n^2 g_{\alpha,n}
		\biggr)
		\right\}
       \nonumber\\
       &&~~~
    + \theta(\tau^F_+-\tau^F_-)\bigg\{
    ( r_- -\alpha^{-1})
		\sum_{n=1}^\infty n g_{-\alpha,n}
		+ r_-  
		\sum_{n=1}^\infty n^2 g_{-\alpha,n}
                \biggr\}\biggr]
  \label{caljx}
\end{eqnarray}
with
\begin{eqnarray}
  r_-\equiv  a(\tau^F_-(x)) \alpha{\cal R}(x) /H,
\end{eqnarray}
$f_\pm = f_\pm(x,x)$ and $g_{\kappa, n} = g_{\kappa,n}(x,x)$ from (\ref{fgdef}), and 
\begin{eqnarray}
  \sum_{n=1}^\infty n g_{\kappa,n}
	&=& \frac{\kappa^2/\pi}{\Omega_- -\Omega_+} \left( \frac{F_{1,\kappa, \Omega_-}}{\Omega_- -\kappa}
     -\frac{F_{1,\kappa,\Omega_+}}{\Omega_+ -\kappa} \right) \\
  \sum_{n=1}^\infty n^2 g_{\kappa,n}
	&=& \frac{\kappa^2/\pi}{\Omega_- -\Omega_+ }\left(\frac{F_{1,\kappa,\Omega_-}}{\Omega_--\kappa}
  -\frac{ F_{1,\kappa,\Omega_+}}{\Omega_+-\kappa}-\frac{2F_{2,\kappa, \Omega_-}}{\Omega_- -2\kappa}
  +\frac{2F_{2,\kappa,\Omega_+}}{\Omega_+-2\kappa}\right)
\end{eqnarray}
where
\begin{equation}
    F_{n,\kappa,w} \equiv  e^{n\kappa (\tau^F_+(x)-\tau^F_-(x))}\,{}_2F_1 \left(n+1, n-\frac{w}{\kappa}; n+1-\frac{w}{\kappa};
		-e^{\kappa (\tau^F_+(x)-\tau^F_-(x))}\right).
\end{equation}

The radiation rate with respect to the proper time of the detector and measured by a global observer in the conformally flat coordinates is obtained by
\begin{eqnarray}
  {dE\over d\tau}= \lim_{\tilde{r}\to\infty}
  \int d\Omega_{(2)}^2 \tilde{r}^2  u^\mu(\tau^F_-) v^\nu(\tau^F_-)  \langle T_{\mu\nu}\rangle^{}_{ren}, \label{dEdtaudef}
\end{eqnarray}
where $\Omega_{(2)}$ denotes the angular variables. The physical radial distance $\tilde{r}$ in the above expression is 
\begin{eqnarray}
  \tilde{r}=a(\eta) \alpha {\cal R}(x)/H 
\end{eqnarray}
in de Sitter spacetime (note that this $\tilde{r}$ is not the radial coordinate $r$ in the static de Sitter coordinates). 
So we have
\begin{eqnarray}
  &&{dE\over d\tau}=
  \lim_{\tilde{r}\to\infty} {2\over 3}{\lambda^2 \alpha \over (4\pi)^2}\int d\Omega_{(2)}^2  {\cal J}(x).
\end{eqnarray}

\subsection{$A=0$}
\textcolor{black}{When the proper acceleration $A$ vanishes, the detector will be inertial and going along a geodesic. We find that the 
energy radiation rate $dE/d\tau$ reduces to zero in this case, 
as was shown in Ref.~\cite{IYZ2013}.}
Indeed, in the late-time limit, $\eta\rightarrow 0$, from Eq.~(\ref{taufpm}), we have
\begin{eqnarray}
a(\tau_-^F)={H\over \alpha^2}{1\over (-HKx+{\cal R}(x))},
\end{eqnarray}
which yields
\begin{eqnarray}
r_--{\alpha}^{-1}={\alpha \over H}a(\tau_-^F(x)){\cal R}(x)-{\alpha}^{-1}={HKx\over \alpha(-HKx+{\cal R}(x))}.
\end{eqnarray}
${\cal J}(x)$ is in proportion to $(r_--{\alpha}^{-1})$ from Eq.~(\ref{caljx}). Then, ${\cal J}(x)$
reduces to zero in the limit of $K=0$, i.e., $A=0$.

\subsection{$A\neq 0$ and $\Omega_\pm \gg \alpha$}

In the case $A\neq 0$ and $\Omega_\pm \gg \alpha$, we may roughly estimate as
${\cal J}(x)\sim g_{\alpha,1}\sim \alpha^2/\Omega^2$. Then we may write the radiation rate as
\begin{eqnarray}
  &&{dE\over d\tau}\sim{\lambda^2\over (4\pi)^2}{\alpha^3\over \Omega^2},
\label{dEdt}
\end{eqnarray}
in the limit $\eta\rightarrow 0$.
This formula is similar to the result in Minkowski
spacetime \cite{LH2006,IOTYZ}, where the radiation rate is expressed as
\begin{eqnarray}
  &&{dE\over d\tau}\sim{\lambda^2\over (4\pi)^2}{A^3\over \Omega^2},
\label{dEdt2}
\end{eqnarray}
except that the proper acceleration $A$ in Minkowski formula is replace by 
$  \alpha=\sqrt{A^2+H^2}$ in Eq.~(\ref{dEdt}). 
So in the conformally flat coordinates for de Sitter spacetime 
a detector going along a straight worldline is accelerating and able to radiate
except the cases with $A=0$. 
Note that each time-slice of the spatially flat chart is a Cauchy surface
of the global de Sitter spacetime. Thus we expect that 
the quantum radiation by the same detector would be similarly nonzero in the global coordinates.

\section{Discussion: Vanishing radiation rates in (1+1)D spacetimes}
\label{discuss2D}

In the detector-field models in (1+1)D Minkowski space, while there still exist nonlocal correlations of the field similar to those in (3+1)D, the radiation rate of a uniformly accelerated detector vanishes due to a special property in (1+1)D field theory as well as the fluctuation-dissipation relation \cite{Raine, HR00}.

In (3+1)D Minkowski spacetime with the minimal oscillator-field coupling, the subtracted two-point correlators of the field up to the frequency integrations are similar to (\ref{subtwo})-(\ref{Fdef}), where the spatial and temporal dependence is concentrated in the factor $e^{i\omega (\tau^F_+(x)-\tau^F_-(x'))}/({\cal R}(x){\cal R}(x'))$ and it's complex conjugate. For a uniformly accelerated detector moving along the worldline $z^\mu(\tau) = (a^{-1}\sinh a\tau, a^{-1}\cosh a\tau, 0,0)$, one has \cite{Lin03, LH2006, HIUY}
\begin{equation}
  \tau^F_+(x) = \frac{1}{a}\ln \frac{a (X-UV + x_\perp^2 +a^{-2})}{2U}, \hspace{.5cm}
	\tau^F_-(x) = -\frac{1}{a}\ln \frac{a (X-UV + x_\perp^2 +a^{-2})}{2V}, 
\end{equation}
where $U\equiv x^0-x^1$, $V\equiv x^0 + x^1$ ($U$, $V>0$ in the F-wedge), $x_\perp^2 = x_2^2 + x_3^2$, and $X\equiv\sqrt{(UV-x_\perp^2+a^{-2})^2+4a^{-2}x_\perp^2}= 2{\cal R}/a$. Thus the factor relevant to the spatial dependence of the radiation is 
\begin{equation}
  \left. \frac{e^{i\omega (\tau^F_+(x)-\tau^F_-(x'))}}{{\cal R}(x){\cal R}(x')}
	\right|_{{\bf R}_1^3} = \frac{4}{a^2 X X'}
	\left[\frac{a^2}{4UV'} (X-UV + x_\perp^2 +a^{-2})(X'-U'V' + x_\perp'^2 +a^{-2}) \right]^{i\omega/a}, \label{eTinM4}
\end{equation}
which gives non-vanishing radiation rate by (\ref{dEdtaudef}) and (\ref{TuvNM}). Note that ${\cal R}^2\langle T_{01} \rangle^{}_{ren}|_{x_\perp=0,{\cal R}\to \infty}\not=0$, namely, the differential radiated energy in $\pm x^1$-direction (parallel to the linear motion of the detector) is not vanishing, due to those terms with the derivatives of the ${\cal R}$ factors \cite{LH2006}.

For the scalar field in (1+1)D Minkowski space, $\xi=0$ for the conformal coupling, which is also the minimal coupling to the scalar curvature \cite{BiD}. 
With the derivative oscillator-field coupling, we simply let $x_\perp=0$, so that $\tau^F_+ \to -a^{-1}\ln aU$ and $\tau^F_- \to a^{-1}\ln aV$, and notice that the retarded Green's function of the field is $G_{ret}(x,x')=\frac{1}{2}\theta(U-U')\theta(V-V')$ and the positive-frequency Wightman function is $D^+(x,x') = (4\pi)^{-1}\ln |\sigma(x,x')|+$constant 
so the $1/{\cal R}$ factors are absent in the field correlators. Then the location-dependent factors in the field correlators reduce to
\begin{equation}
  \left.e^{i\omega (\tau^F_+(x)-\tau^F_-(x'))}\right|_{{\bf R}_1^1} = 
	\left[ a^2 UV' \right]^{-i\omega/a} =\left[ a^2 (x^0-x^1)(x'^0+x'^1) \right]^{-i\omega/a}.
\label{UVVU}
\end{equation}
Inserting this into the subtracted two-point correlator (\ref{subtwo}) and symmetrize it, the derivatives in (\ref{TuvNM}) immediately give $\langle T_{\mu\nu} \rangle^{}_{ren}|_{\xi=0}=0$ for all $\mu,\nu = 0,1$. For the cases with $\xi\not= 0$, it is still straightforward to get $\langle T_{01} \rangle^{}_{ren}=0$, though $\langle T_{00} \rangle^{}_{ren}$ and $\langle T_{11} \rangle^{}_{ren}$ may not vanish. Thus one concludes that the radiation rate is zero for a uniformly accelerated, derivative-coupling Unruh-DeWitt detector in ${\bf R}_1^1$,
though the nonlocal correlations of the field do exist.

One may wonder if the conformal symmetry in (1+1)D spacetime caused this result. Indeed, conformal symmetry guarantees a traceless stress-energy tensor classically, and one expects that $\langle g^{\mu\nu}T_{\mu\nu}\rangle = -4\langle T_{UV} \rangle$ would be vanishing or equal to the trace anomaly for the models with conformal field-curvature coupling ($\xi=0$).  
However, even if $\langle T_{UV} \rangle\not=0$, the symmetric property $\langle T_{UV}\rangle_{ren} = \langle T_{VU}\rangle_{ren}$ is sufficient to make $\langle T_{01}\rangle_{ren}=\langle T_{VV}\rangle_{ren} -\langle T_{UU}\rangle_{ren} +\langle T_{UV}\rangle_{ren} -\langle T_{VU}\rangle_{ren}$ independent of $\langle T_{UV} \rangle_{ren}$. Thus conformal symmetry is irrelevant to the vanishing energy radiation rate here in (1+1)D Minkowski spacetime.

The direct cause of the vanishing result is that in (1+1)D a massless field can always be split into the right-mover and left-mover, which depend only on $U$ and $V$, respectively (e.g. $\hat{\phi}(x) = \hat{\phi}^R(U)+ \hat{\phi}^L(V)$). This property insists even in the linearly interacting models, and implies that the $U$- and $V$-derivatives of the field will select the right-mover and left-mover, respectively ($\partial_U \hat{\phi}(x) = \partial_U \hat{\phi}^R(U)$ and $\partial_V \hat{\phi}(x) = \partial_V \hat{\phi}^L(V)$). Suppose the detector is uniformly accelerated in the R-wedge. Then in the F-wedge the retarded field $\hat{\phi}_{inh}(x) = \hat{\phi}_{inh}^L(V)$ will be purely left-moving, implying $\langle \partial_U\hat{\phi}(x) \partial_{U'}\hat{\phi}(x')\rangle_{ren}  = \langle \partial_U\hat{\phi}(x) \partial_{U'}\hat{\phi}(x')\rangle -\langle \partial_U\hat{\phi}_{h}^R(U) \partial_{U'}\hat{\phi}_h^R(U')\rangle = 0$ and so $\langle T_{UU}(x) \rangle_{ren}$ must vanish in the F-wedge.
On the other hand, in the equilibrium conditions at late times, the fluctuation-dissipation relation like (\ref{FDR}) offers the necessary cancellation in section \ref{2pCorr} to make the correlations of the left-movers $\langle \partial_V\hat{\phi}(x) \partial_{V'}\hat{\phi}(x')\rangle$ in the presence of the detector at any two points $x$, $x'$ in the F-wedge getting the same value of the ones for the free field, and so the left-moving renormalized energy flux $\langle T_{VV}(x)\rangle_{ren}$ is vanishing in the F-wedge, too,
implying $\langle T_{01}\rangle_{ren}=\langle T_{VV}\rangle_{ren} -\langle T_{UU}\rangle_{ren}=0$ \cite{HR00, HIUY}.
As we mentioned, in (3+1)D Minkowski spacetime, those terms independent of the derivatives of ${\cal R}$ in the renormalized stress energy tensor also vanish at $x_\perp^2=0$, or $\theta=0$, $\pi$, because of the same property.

Now in the model considered in this paper with the minimal oscillator-field coupling and the conformal field-curvature coupling in (3+1)D de Sitter spacetime, (\ref{taufpm}) yields
\begin{equation}
  \left.\frac{e^{i\omega (\tau^F_+(x)-\tau^F_-(x'))}}{a(\eta){\cal R}(\bar x)a(\eta'){\cal R}(\bar x')}\right|_{dS_4} = 
  \frac{1}{a(\bar x^0){\cal R}a(\bar x'^0){\cal R}'}\left[ \frac{{\cal R}'-\bar x'^0-HK\bar x'^1}{{\cal R} +\bar x^0 +HK\bar x^1} 
	\right]^{i\omega/\alpha},
\end{equation}
which looks different from (\ref{eTinM4}) since the worldline of the uniformly accelerated detector here is a straight line rather than a hyperbola in the conformally flat coordinates. In (1+1)D de Sitter spacetime,
  as the retarded Green's function for $\xi=0$ is similar to the one in (1+1)D Minkowski
  spacetime (see e.g., \cite{MTY}),
the location-dependent factors in the late-time field correlators in the presence of a uniformly accelerated detector derivatively coupled to the scalar field reduce to 
\begin{eqnarray}
  \left.e^{i\omega (\tau^F_+(x)-\tau^F_-(x'))}\right|_{dS_2} &=&
  \left( \frac{|\bar x'{}^{1}+HK\bar x'{}^0|-\bar x'{}^0-HK\bar x'{}^1}
    {|\bar x^1+HK\bar x^0|+\bar x^0+HK\bar x^1} \right)^{i\omega/\alpha}
  \nonumber\\
  &=& \left\{
  \begin{array}{cc}
    ~~~~\displaystyle{
      \left( \frac{(HK-1)\bar{U}'}{(HK+1)\bar{V}} \right)^{i\omega/\alpha}}
    ~~& ~~{\rm for}~~
      \bar x^1+HK\bar x^0>0, ~\bar x'{}^1+HK\bar x'{}^0>0,
      \\
     ~~~~\displaystyle{
            \left( \frac{(HK+1)\bar{V}'}{(HK-1)\bar{U}} \right)^{i\omega/\alpha}}
     ~~& ~~{\rm for}~~
           \bar x^1+HK\bar x^0<0, ~\bar x'{}^1+HK\bar x'{}^0<0,
  \end{array}
    \right.
  \end{eqnarray}
  after taking $x_\perp^2=0$.
Here $\bar{U} = \bar{x}^0-\bar{x}^1$, $\bar{V} = \bar{x}^0+\bar{x}^1$,
$\bar{U}'=\bar{x}'^0 - \bar{x}'^1$, and $\bar{V}'=\bar{x}'^0 + \bar{x}'^1$.
It is straightforward to see that $\langle T_{\bar{0}\bar{1}}\rangle_{ren} |_{\xi=0} = 0$ again, due to the similar splitting of the left- and right-movers as well as the fluctuation-dissipation relation
\footnote{While the case with the \textcolor{black}{non-minimal/conformal coupling $\xi\not= 0$ in (1+1)D could be interesting,
the non-minimal/conformal} coupling term to the curvature in general produces an effective mass of the field in de Sitter spacetime, where more complicated analysis is needed.}.

\section{Conclusion}
\label{conclude}

We have investigated the late-time quantum radiation emitted by
a uniformly accelerated detector in de Sitter spacetime. 
Our approximated result Eq.~(\ref{dEdt}) 
is a simple generalization from the radiation rate at large acceleration in Minkowski
spacetime 
(\ref{dEdt2}), 
with the proper acceleration $A$ in Minkowski spacetime 
replaced by $\alpha = \sqrt{A^2+H^2}$ in de Sitter spacetime. 
Indeed, a uniformly accelerated detector in de Sitter spacetime is thermally
excited at the temperature $T=\sqrt{A^2+H^2}/2\pi$,
which reflects the thermal properties of both the Unruh effect and the
Gibbons-Hawking effect \cite{UdeS}.

The Bunch-Davies vacuum state $|0\rangle_{\rm BD}$ in de Sitter spacetime
can be expressed in terms of the states defined separately in the two static charts as \cite{TS94,HY}
\begin{eqnarray}
  |0  \rangle_{\rm BD}={\cal N} \prod_{j} \sum_{n_j=0}^\infty e^{-\pi \omega n_j} |n_j\rangle_R
  \otimes|n_j\rangle_L,
\end{eqnarray}
where ${\cal N}$ is a normalization constant, 
and $|n_j\rangle_R$ and $|n_j\rangle_L$ are the $n_j$-th excited states of a
pair of Fourier modes constructed in the R- and L-regions in the static coordinates.
Here each $j$ denotes a Fourier mode of $(\omega$, $\ell$, $m)$, where $\ell$
and $m$ are the degree and the order for the spherical harmonics. 
This expression 
is similar to 
the one 
for the Minkowski vacuum state in terms of the right- and left-Rindler modes 
defined in the R- and L- wedges, respectively.
In the case of Minkowski spacetime, the late-time quantum radiation 
emitted by a uniformly accelerated detector is 
originated from the non-local correlation of the vacuum state
of the field, which can be traced back to the 
entanglements between the right- and left-Rindler modes 
\cite{IOTYZ,ITUY,HIUY}. 
The results in the present paper shows 
that the late-time quantum radiation produced 
by a uniformly accelerated detector in de Sitter space 
similarly comes from the non-local correlation of the vacuum state
which can be traced back to the entanglements between the modes in the R- and L-regions 
in de Sitter spacetime. 

\textcolor{black}{
In Minkowski spacetime, the causal structure for a uniformly accelerated detector looks quite different from the one for a free-falling detector. There exist horizons and late-time quantum radiation for the former, but neither for the latter. Thus one is tempted to say that the presence of the horizons plays an important role in producing the nonvanishing late-time quantum radiation. Nevertheless, an inertial Unruh-DeWitt detector moving along a geodesic in de Sitter spacetime shares the same cosmological horizon with a family of the uniformly accelerated detectors (see Figure 1), and the late-time quantum radiation is vanishing for the inertial detector.  
Thus the presence of the horizons does not implies non-vanishing quantum radiation at late times, though it is essential for the nonlocal correlations of the field modes and the thermal property that the detector experiences in vacuum. 
}

\begin{acknowledgments}
We thank A. Higuchi, B.~L.~Hu, S. Iso, S.~Kanno, Y.~Nambu, M.~Sasaki, J.~Soda, and
K. Ueda for useful communications related to the topic in the present paper. 
This work is supported by MEXT/JSPS KAKENHI Grant Numbers 15H05895, 
17K05444, and 17H06359, the MOST of Taiwan 
Grant No. 106-2112-M-018-002-MY3, and in part by the 
National Center for Theoretical Sciences, Taiwan. 
\end{acknowledgments}




\begin{thebibliography}{99}

\bibitem{Unruh}
 W. G. Unruh, Phys. Rev. D {\bf 14}, 870 (1976).

\bibitem{ref-4}
S.~Takagi, Prog.~Theor.~Phys.~Sppl. {\bf88}, 1 (1986).

\bibitem{ChenTajima}
 P.~Chen, T.~Tajima, Phys. Rev. Lett. {\bf 83}, 256 (1999).

\bibitem{Schutzhold}
 R. Sch\"utzhold, G. Schaller, D.Habs, Phys. Rev. Lett. {\bf 97}, 121302 (2006).

\bibitem{Schutzhold2}
 R. Sch\"utzhold, G. Schaller, D.Habs, Phys. Rev. Lett. {\bf 100}, 091301 (2008).

\bibitem{ELI}
  P.G.~Thirolf, D.~Habs, A.~Henig, D.~Jung, D.~Kiefer, C.~Lang, J.~Schreiber1, C.~Maia, G.~Schalle, R.~Sch\"utzhold,
  and T.~Tajima, Eur.\ Phys.\ J.\ D {\bf 55}, 379 (2009).

\bibitem{IYZ}
 S.~Iso, Y.~Yamamoto, S.~Zhang, Phys. Rev. D {\bf 84}, 025005 (2011).

\bibitem{Raine}
 D. J. Raine, D. W. Sciama, P. G. Grove, Proc. R. Soc. Lond. {\bf A435}, 205 (1991).

\bibitem{Raval}
 A. Raval, B. L. Hu, J. Anglin, Phys. Rev. D {\bf 53}, 7003 (1996).

\bibitem{Ford}
  G. W. Ford, R. F. O'Connell, Phys. Lett. {\bf A350}, 17 (2006).

\bibitem{Lin16}
  S.-Y. Lin, in {\it Proceedings of the MG14 Meeting on General Relativity},
  edited by M. Bianchi, R. T. Jantzen, and R. Ruffini (World Scientific 2017), arXiv:1601.07006

\bibitem{Lin17} S.-Y. Lin, 
  {\it Quantum radiation by an Unruh-DeWitt detector in oscillatory motion},
  JHEP {\bf 11 (2017)} 102. 

\bibitem{LH2006}
  S.-Y. Lin and B.L.Hu, Phys. Rev. D {\bf 73}, 124018 (2006);
	Found. Phys. {\bf 37}, 480 (2007).	

\bibitem{HIUY}
 A. Higuchi, S. Iso, K. Ueda, K. Yamamoto, Phys. Rev. D {\bf 96}, 083531 (2017).

\bibitem{ITUY}
 S. Iso, R. Tatsukawa, K. Ueda, K. Yamamoto, Phys. Rev. D {\bf 96}, 045001 (2017). 

\bibitem{IOTYZ}
 S. Iso, N. Oshita, R. Tatsukawa, K. Yamamoto, S. Zhang, Phys. Rev. D {\bf 95}, 023512 (2017).

\bibitem{BiD} 
 N. D. Birrell and P. C. W. Davies, Quantum fields in curved space (Cambridge University Press 1982)

\bibitem{UnruhWald}
 W. G. Unruh, R. M. Wald, Phys. Rev. D {\bf 29}, 1047 (1984).

\bibitem{Higuchi}
 L. C. B. Crispino, A. Higuchi, G. E. A. Matsas, Rev. Mod. Phys. {\bf 80}, 787 (2008).

\bibitem{OYZ15}
 N.~Oshita, K.~Yamamoto, S.~Zhang, Phys. Rev. D {\bf 92}, 045027 (2015).

\bibitem{OYZ16}
 N.~Oshita, K.~Yamamoto, S.~Zhang, Phys. Rev. D {\bf 93}, 085016 (2016).

\bibitem{KukitaNambu}
S. Kukita, Y. Nambu,  Class. Quant. Grav. {\bf 34}, 235010 (2017). 

\bibitem{Rotondo}
  M. Rotondo, Y. Nambu, Universe {\bf 3}, 71 (2017). 

\bibitem{KukitaNambu2}
  S. Kukita, Y. Nambu,  Entropy {\bf 19},  449  (2017). 
  
\bibitem{MatsumuraNambu}
  A. Matsumura, Y. Nambu, Phys. Rev. D {\bf 98}, 025004 (2018). 

\bibitem{MaldacenaPimentel}
  J. Maldacena, G. L. Pimentel, JHEP {\bf 02 (2013)} 038.

\bibitem{Kanno}
  S. Kanno, J. P. Shock, J. Soda, JCAP {\bf 03 (2015)} 015.

\bibitem{KST}
  S. Kanno, M. Sasaki, T. Tanaka, JHEP {\bf 03 (2017)} 068.

\bibitem{GH}
  G. Gibbons, S. Hawking Phys. Rev. D {\bf 15} 2738 (1977).

\bibitem{BD}
   T. S. Bunch and P. C. Davies, Proc. R. Soc. London {\bf A 360}, 117 (1978).

\bibitem{IYZ2013}
 S.~Iso, K.~Yamamoto, S.~Zhang, Prog. Theor. Exp. Phys. {\bf 2013}, 063B01.

\bibitem{MTY}
  T. Murata, K. Tsunoda, K. Yamamoto, Int. J. Mod. Phys. A {\bf 16}
  2841 (2001). 

\bibitem{UdeS}
  R. Casadio, S. Chiodini, A. Orlandi, G. Acquaviva, R. Di Criscienzo, L. Vanzo,
  Mod. Phys. Lett. A {\bf 26}, 2149 (2011). 
	
\bibitem{Burko}
L. M. Burko, A. I. Harte, and E. Poisson, Phys. Rev. D {\bf 65} 124006 (2002) 

\bibitem{Akhmedov2010}
  E. T. Akhmedov, A. R. A. Sadofyev, Phys. Rev. D {\bf 82}, 044035 (2010).

\bibitem{Blaga2015}
  R. Blaga, Mod. Phys. Lett. A {\bf 30}, 1550062 (2015).

\bibitem{BB16}	
  R. Blaga and S. Busuioc, Eur. Phys. J. C {\bf 76}, 500 (2016).

\bibitem{NSY06} H. Nomura, M. Sasaki, K. Yamamoto, JCAP {\bf 11 (2006)} 013.

\bibitem{KNY11} R. Kimura, G. Nakamura, K. Yamamoto, Phys. Rev. D {\bf 83}, 045015 (2011).

\bibitem{BL83} J. S. Bell and J. M. Leinaas, 
    Nucl. Phys. {\bf B 212}, 131 (1983).

\bibitem{BL87} J. S. Bell and J. M. Leinaas, 
    Nucl. Phys. {\bf B 284}, 488 (1987).

\bibitem{Un98} W. G. Unruh, 
in {\it Monterey Workshop on Quantum Aspects of Beam Physics}, edited by Pisin Chen
     (World Scientific, Singapore, 1998) [hep-th/9804158];  
     Phys. Rep. {\bf 307}, 163 (1998). 
     
\bibitem{AS07} E. T. Akhmedov and D. Singleton, 
       Int. J. Mod. Phys. {\bf A 22}, 4797 (2007); 
			 E. T. Akhmedov and D. Singleton, 
       Pisma Zh. Eksp. Teor. Fiz. {\bf 86}, 702 (2007) 
			 [JETP Lett. {\bf 86}, 615 (2007)].

\bibitem{Akhmedov2012}
  E. T. Akhmedov, P. V. Buividovich, D. A. Singleton, Phys. Atom Nucl. {\bf 75}, 525 (2012).
 
\bibitem{JohnsonHu}
  P. R. Johnson and B. L. Hu, Phys. Rev. D {\bf 65}, 065015 (2002);
  Found. Phys . {\bf 35} 1117 (2005)

\bibitem{OLMH12} D. C. M. Ostapchuk, S.-Y. Lin, R. B. Mann, and B. L. Hu,
JHEP {\bf 07 (2012)} 072. 

\bibitem{Rohrlich}
  F. Rohrlich, {\it Classical Cahrged Prticle } (Addison-Wesley, Reading MA, 1965).
	
\bibitem{HR00} B. L. Hu and A. Raval, 
			in {\it Proceedings of the Capri Workshop on Quantum Aspect of Beam Physics, Oct. 2000}, edited by Pisin Chen
			(World Scientific, Singapore, 2001), quant-ph/0012135.
  
\bibitem{Lin03} S.-Y. Lin, 
  Phys. Rev. D {\bf 68}, 104019 (2003).

\bibitem{TS94}
  T. Tanaka, M. Sasaki, Phys. Rev. D {\bf 50}, 6444 (1994).

\bibitem{HY}
  A. Higuchi, K. Yamamoto, Phys. Rev. D {\bf 98}, 065014 (2018).



  
\end{thebibliography}
\end{document}